\documentclass[12pt, notitlepage,aps,preprintnumbers,superscriptaddress,nofootinbib,aps,prd,floatfix]{revtex4-2}
\usepackage{float}
\usepackage{physics,slashed}
\usepackage{amsfonts,amsmath,amssymb}
\usepackage{gensymb}
\usepackage{mathrsfs}
\usepackage{bm}

\usepackage{epsfig}
\usepackage{graphicx}   
\usepackage{subfigure}  
\usepackage{verbatim}   
\usepackage{color}      
\usepackage{hyperref}   
\usepackage{url}
\definecolor{nicered}{rgb}{0.5,0.,0.}
\definecolor{nicegreen}{rgb}{0.,0.5,0.}
\definecolor{niceblue}{rgb}{0.,0.,0.5}
\hypersetup{
    colorlinks=true,      
    citecolor=nicegreen,  
    linkcolor=niceblue,      
    urlcolor=niceblue,    
    linktocpage=true,     
    pagecolor=nicered     
}
\usepackage{array}
\usepackage{dcolumn}
\usepackage{multirow}
\usepackage{cprotect}
\usepackage{framed}
\usepackage{setspace}
\usepackage[table]{xcolor}
\usepackage{tabularx}

\usepackage{graphicx}
\usepackage{subcaption}
\usepackage{feynmp-auto} 

\usepackage{tikz-feynman}
\tikzfeynmanset{compat=1.1.0}

\usepackage{feynmp-auto}
\usepackage{capt-of}

\usepackage{tikzsymbols}

\renewcommand{\arraystretch}{1.4} 

\def\beq{\begin{equation}}
\def\eeq{\end{equation}}

\def\mO{\mathcal{O}}

\begin{document}
\preprint{PITT-PACC-2406}
\title{Top Quark Electroweak Dipole Moment\\ at a High Energy Muon Collider}

\author{Tao Han}
\affiliation{Pittsburgh Particle Physics, Astrophysics, and Cosmology Center, Department of Physics and Astronomy, University of Pittsburgh, Pittsburgh, PA 15206, USA\looseness=-1}
\author{Da Liu}
\affiliation{Pittsburgh Particle Physics, Astrophysics, and Cosmology Center, Department of Physics and Astronomy, University of Pittsburgh, Pittsburgh, PA 15206, USA\looseness=-1}
\author{Si Wang}
\affiliation{Pittsburgh Particle Physics, Astrophysics, and Cosmology Center, Department of Physics and Astronomy, University of Pittsburgh, Pittsburgh, PA 15206, USA\looseness=-1}


\begin{abstract}
We study the sensitivity reach to probe the electroweak dipole operators associated with a top quark at a multi-TeV lepton collider. Studying the electroweak dipole operators is strongly motivated by precision physics. The operators exhibit unique chiral  structure and can be enhanced with respect to others in theories beyond the Standard Model. We illustrate this point in a strongly coupled composite Higgs model. 
We find that a high energy $\mu^+\mu^-$ collider may offer unique opportunity to probe the electroweak dipole operators beyond the coverage by the LHC and future hadron colliders. The significant sensitivity is achieved by the leading channel $\mu^+ \mu^- \to t \bar t$ near the threshold, and substantially improved by the novel channel $\mu^+ \mu^- \to t \bar t h$ at high energies. We may be able to reach a new physics scale well above the collider energy for moderate couplings of the Wilson coefficients.
Our main conclusions are applicable to future $e^+e^-$ colliders. 
\end{abstract}
\maketitle
\tableofcontents

\newpage
\section{Introduction}
\label{sec:intro}

The outstanding experiments of ATLAS and CMS at the CERN Large Hadron Collider (LHC) made the milestone discovery of the Higgs boson \cite{ATLAS:2012yve,CMS:2012qbp}, and the Standard Model (SM) of the strong and electroweak interactions has been established to a high precision \cite{ParticleDataGroup:2024cfk}. 
The SM-like Higgs boson brings in new questions regarding the underlying mechanism for the electroweak symmetry breaking. The top quark, as the heaviest particle in the SM with the strongest interaction to the Higgs boson, may hold the key to understanding the nature beyond the electroweak scale. 

Electromagnetic dipole moments of elementary particles have been an important subject to study through the particle physics history. It helped to reveal the composite nature of the proton \cite{PhysRev.46.157}, to establish the foundation of Quantum Electro-dynamics (QED) \cite{PhysRev.73.416} through the prediction of electron $(g-2)$, and to explore new physics effects with precision measurements at low energies \cite{ACME:2018yjb, ACME:2013pal, Cairncross:2017fip, Muong-2:2006rrc,Muong-2:2021ojo,Muong-2:2023cdq}. Marching into the higher energy regime at colliders, the electroweak dipole moments of heavy fermions will continue to be scrutinized in search for new physics at the LHC \cite{Baur:2004uw,Ellis:2008zy} and future $e^+e^-$ colliders \cite{Asner:2013hla,Billur:2017yds}. 

When not knowing the structure of the underlying physics, the Effective Field Theory (EFT) provides an adequate formulation to parameterize the ignorance. 
The EFT Lagrangian may be expressed by a power expansion in terms of an energy scale $\Lambda$ at which some new physics sets in.
Beyond the SM at dimension-6, it is given by 
\beq
\mathscr{L}=\mathscr{L}_{\rm SM} + \sum_{i} (\frac{c_i}{\Lambda^2} \mO_i + h.c.) . 
\label{eq:SMEFT}
\eeq
where $\mO_i$ are dim-6 operators in terms of the SM fields and satisfying the SM gauge symmetries \cite{Leung:1984ni,Buchmuller:1985jz}.
This is often referred as the Standard Model effective field theory (SMEFT) \cite{Grzadkowski:2010es}.
The Wilson coefficients $c_i$ are a prior unknown. In terms of an underlying theory, they may consist of factors of SM gauge couplings, Yukawa couplings and some loop factors dependent on the integrated-out UV physics \cite{Henning:2014wua,Giudice:2007fh}. 
In a weakly-coupled underlying theory at the appropriate new physics scale 
$\Lambda$, the Wilson coefficients $c_i$ are usually less than unity and are products of the SM couplings.
For a strongly-coupled underlying theory, they can be enhanced by factors of the typical strong coupling $g_*$ dependent on whether the operators preserve the global symmetry of the strong sector or not.
For example, in the framework of strongly-interacting light Higgs (SILH) \cite{Giudice:2007fh}, the modifications to the Higgs couplings to SM gauge bosons and fermions from the new physics can be enhanced by $g_*^2$ compared with weakly coupled theory like SUSY. Note that the new coupling $g_*$ is in the range of $\sim 1- 4\pi$, and in the large-$N$ limit of QCD-like theory~\cite{Witten:1979kh}, it scales like $4 \pi / \sqrt{N}$. 
%
As another concrete example as in Ref.~\cite{Liu:2016idz}, a model with the scenario of strongly coupled multi-polar interaction has been proposed. In such a scenario, dubbed as ``Remedios"  \cite{Liu:2016idz}, the SM electroweak (transverse)  gauge bosons are part of the strong dynamics, in addition to the Higgs doublet. The Wilson coefficients of the operators involving the field strength of SM gauge bosons 
$W^{a}_{\mu\nu}$ and $ B_{\mu\nu}$ can be potentially enhanced by a strong coupling $g_*$. 

 With the recent significant efforts of the feasibility studies of future colliders reaching a partonic center-of-mass energy of 10 TeV, the expectation is high to improve our understanding of particle physics at shorter distances \cite{Narain:2022qud}. It is natural to extend the existing studies for the electroweak dipole operators \cite{Janot:2015yza,Amjad:2013tlv, Koksal:2016wlh,Billur:2017yds} to consider the sensitivity at the high energy lepton colliders, such as CLIC \cite{CLICdp:2018cto} and a muon collider \cite{Delahaye:2019omf,Aime:2022flm,Black:2022cth,InternationalMuonCollider:2024jyv}. 
 In this paper, we consider the electroweak dipole operators involving the top quark and a Higgs boson. The primary motivation is that the dipole operators as well as the Higgs Yukawa couplings involve the chirality-flip interactions and thus potentially more sensitive to new physics associated with heavy fermions. 
We find that the major contributions come from the leading processes 
 \begin{equation}
 \mu^+ \mu^- \to t \bar t\quad {\rm and}\quad  \mu^+ \mu^- \to t \bar t h.
 \end{equation}
 The $t\bar th$ associate production becomes increasingly significant due to the enhanced interference terms with the new couplings. 
 In high energy $\mu^+ \mu^-$ collisions, the rate for the vector boson fusion (VBF) increases with the collision energy logarithmically \cite{Costantini:2020stv,Han:2020uid}, and some work existed to utilize the VBF at a muon collider to probe the top-quark Yukawa coupling \cite{Chen:2022yiu,Liu:2023yrb,Barger:2023wbg,Cassidy:2023lwd}.
%
We include the VBF contributions
$\mu^+ \mu^- \to \mu^+ \mu^- t \bar t$ and $\mu^+ \mu^- \to \nu \bar \nu t \bar t$ in considering the top-quark electroweak dipole interactions.

 The rest of the paper is organized as follows. In Sec.~\ref{sec:theo}, we introduce the electroweak dipole moment operators in terms of the effective field theory (EFT) at dimension-6, and provide theoretical motivation illustrated by a strongly-interacting composite Higgs model. We then compile the constraints from the current existing measurements and from future collider expectations on the top-quark magnetic and electric dipole moments, and translate them to the lower bounds on the Wilson coefficients and the mass scale of our current interests.
In Sec.~\ref{sec:sen}, we calculate the production cross sections for the processes $\mu^+ \mu^- \to t \bar t X$, and including the new physics contributions. We perform a detailed signal-background analysis and derive the $2\sigma$ sensitivity bounds on the model parameters at a few benchmark energies, and extend the results to the energy range to $\sqrt s=10$ TeV. The summary and conclusions are given in Sec.~\ref{sec:sum}.  


\section{Theoretical Formulation}
\label{sec:theo}

\subsection{Electroweak dipole operators in SMEFT and Remedios}

We adopt the effective field theory (EFT) approach in parameterizing the physics beyond the SM. 
In SMEFT at dim-6 in the Warsaw basis \cite{Grzadkowski:2010es}, the electroweak dipole operators in Eq.~(\ref{eq:SMEFT}) are of the general form
%
\beq
\begin{split}
\mO_{uB} &= \bar{Q}_L \sigma^{\mu\nu}u_R \tilde{H} B_{\mu\nu}, \qquad \mO_{uW} =  \bar{Q}_L \sigma^{\mu\nu}u_R  \sigma^a\tilde{H} W^a_{\mu\nu}, \\
\mO_{dB} &=  \bar{Q}_L \sigma^{\mu\nu}d_R H B_{\mu\nu}, \qquad \mO_{dW} = \bar{Q}_L \sigma^{\mu\nu}d_R  \sigma^aH W^a_{\mu\nu},\\
\mO_{lB} &= \bar{L}_L \sigma^{\mu\nu} l_R H B_{\mu\nu}, \qquad \mO_{lW} =  \bar{L}_L \sigma^{\mu\nu} l_R \sigma^a H W^a_{\mu\nu} .
\end{split}
\label{eq:O}
\eeq
Here $W^a_{\mu\nu}$ and $B_{\mu\nu}$ are the $SU(2)$ and $U(1)$ field strength tensors, $\sigma^{\mu\nu}=i(\gamma^\mu \gamma^\nu-\gamma^\nu \gamma^\mu)/2$, and $\sigma^a$ the Pauli matrices. $Q_L$ ($L_L$) is the left-handed quark (lepton) doublet and $u_R\ (d_R)$ is the right-handed up-type (down-type) quark singlet. $l_R$ is the right-handed charged lepton, and $H \ (\tilde{H})$ is the Higgs doublet. 
 
In this paper, we focus on the electroweak dipole operators involving the top quark. The effective Lagrangian under consideration includes two terms in Eq.~(\ref{eq:O}), $\mO_{tB}$ and $\mO_{tW}$, and they takes the form 
%
\begin{eqnarray}
\mathcal{L}'=\frac{c_{tB}}{\Lambda^2}\bar Q_L \sigma^{\mu\nu}t_R \Tilde{H}B_{\mu\nu} +\frac{c_{tW}}{\Lambda^2}\bar Q_L \sigma^{\mu\nu} \sigma^a t_R \Tilde{H}W_{\mu\nu}^a
+ h.c.
\label{eq:operator}
\end{eqnarray} 
The Wilson coefficients $c_{tB}$ and $c_{tW}$ are a priori unknown and can be large or small depending on the underlying  UV physics. 
$\Lambda$ is the physical scale at which new physics sets in. Those are the theory parameters that we will set out to probe at future colliders. After spontaneous electroweak symmetry breaking,
the Wilson Coefficients of the physical gauge bosons,  $\gamma$ and $Z$, are given by
\begin{eqnarray}
c_{t\gamma}=c_w c_{t B} + s_w c_{tW}, \quad 
c_{tZ}=s_w c_{t B} - c_w c_{tW},  
\label{eq:mix}
\end{eqnarray}
where $c_w\ (s_w)$ are the cosine (sine) of the weak mixing angle ($s_w^2\approx 0.231$). 

Our interest in the electroweak dipole moment operators is particularly motivated by their potentially large Wilson coefficients studied 
in Ref.~\cite{Liu:2016idz}, where a scenario of strongly coupled multi-polar interaction has been proposed. In such a scenario, the SM transverse gauge bosons are part of the strong dynamics.  We will have the following new power-counting rules for the gauge bosons: the field strengths of the gauge bosons and their derivatives will be associated with a strong coupling $g_*$ as long as the interactions preserve the global symmetries of the strong sector. The normal SM gauge interactions are realized by changing the partial derivative to covariant derivative  $\partial_\mu \rightarrow D_\mu = \partial_\mu -i  g W_\mu^a \sigma^a/2 - Y g^\prime B_\mu $ with $g, g^\prime$ as the corresponding $SU(2)_L, U(1)_Y$ gauge couplings, which can be seen as the deformation of the original symmetry.
If we also consider  further the scenarios that both transverse gauge bosons and Higgs bosons are part of the strong dynamics, depending on the symmetry of the strong sector, there are two representative benchmark scenarios: 
\begin{itemize}
\item \text{Remedios + MCHM} (Minimal Composite Higgs Model~\cite{Contino:2003ve,Agashe:2004rs}): the symmetry breaking of the strong sector will be $SO(5)\times \widetilde{SU(2)} \times U(1)_X \rightarrow SO(4)\times \widetilde{SU(2)} \times U(1)_X$, where another global symmetry $\widetilde{SU(2)}$ is needed to stabilize the Higgs potential. 
\item \text{Remedios} +$ ISO(4)$: the symmetry breaking of the strong sector will be $ISO(4)\times  U(1)_X \rightarrow SO(4) \times U(1)_X$, where the $ISO(4)$ is the non-compact group $SO(4)\rtimes T^4$.
\end{itemize}
Before fully accessing the new physics scale $\Lambda$, the new physics effects may reveal themselves as large composite operators. 
Such an example was presented in a scenario of SILH (Strongly Interacting Light Higgs) 
\cite{Giudice:2007fh}
with enhanced effects by the strong coupling consist $g_*$. 
%
Together with the scenarios of our current interest, the corresponding power-counting rules for the size of the Wilson coefficients $c_i$ are summarized in Table~\ref{tab:wilsoncoeff}. Following the arguments in the Naive Dimensional Analysis (NDA) \cite{Cohen:1997rt} for a strong coupling $g_*\sim 4\pi$, we may expect the Wilson coefficients associated with the top quark $c_{t B},\ c_{t W}$ to reach as large as the order of $4\pi$ at an appropriate scale $\Lambda$. 



 \begin{table}[tb]
\centering
\renewcommand{\arraystretch}{1.5}
\begin{tabular}{|c|c|c|c|c|c|c|c|c|c|c|c|c|}
\hline 
Model   & $\mO_{lB}$ & $\mO_{lW} $& $\mO_{qB}$ & $\mO_{qW} $\\
\hline
{\footnotesize SILH }~\cite{Giudice:2007fh}   &$\frac{g_*^2}{16\pi^2} y_l g^\prime$ & $\frac{g_*^2}{16\pi^2} y_l g$   &$\frac{g_*^2}{16\pi^2} y_q g^\prime$ & $\frac{g_*^2}{16\pi^2} y_q g$  \\
{\footnotesize  Remedios+MCHM  }~\cite{Liu:2016idz}  & $ y_l g_*$ & $y_l g$ & $y_q g_*$ & $y_q g$ \\
{\footnotesize  Remedios+$ISO(4)$  }~\cite{Liu:2016idz}  & $y_l g_*$ & $y_lg_*$  & $y_q g_*$ & $y_qg_*$ \\
\hline
\end{tabular}
\caption{Power counting of the size of the Wilson coefficients $c_i$ as in Eq.~(\ref{eq:operator}) in the representative scenarios, where $g_*$ denotes the coupling in the strongly interacting sector and $q = u, d$. 
}
\label{tab:wilsoncoeff}
\end{table}%


A special remark is in order regarding the test sensitivity of the dipole operators in Eq.~({\ref{eq:operator}). 
We note that these operators flip the fermion chirality, and thus are less sensitive to light fermions due to the mass insertion at the leading order. A similar argument would apply to the top quark at high energies when $m_t/E \ll 1$. We will demonstrate this point in our analyses later. 

\subsection{Current constraints on top-quark dipole operators}

In QED, the effective $t\bar t \gamma$ Lagrangian can be written as 
\begin{eqnarray}
\label{eq:avaadef}
\mathcal{L}_{t\bar t \gamma}=e \bar t \left(Q_t\gamma_\mu A^\mu + \frac{1}{4m_t}\sigma_{\mu\nu}F^{\mu\nu}
(a_V+i a_A \gamma_5)\right)t , 
\end{eqnarray}
where $A^\mu$ and $F^{\mu\nu}$ are the photon field and the field-strength, respectively and. $a_V$ and $a_A$ are related to the top-quark anomalous magnetic and electric dipole moments $a_t$ and $d_t$ as 
\begin{equation}
   a_V=Q_t a_t,\quad a_A=\frac{2m_t}{e}d_t , 
\end{equation} 
with $Q_t$ the top-quark electric charge ($+2/3$), $m_t$ the top quark mass ($\approx 173$ GeV), and $e$ the elementary charge $(\approx 0.3)$, with 
$2m_t /e \approx 1.754 \times  10^{16}\ ({\rm e\cdot  cm})^{-1}$. 
In reverse, the electromagnetic dipole moments are given as $a_t=3/2 a_V, d_t=0.57\times 10^{-16} a_A\ ({\rm e}\cdot {\rm cm})$ \cite{Bouzas:2012av}. 
In the SM, the top-quark $g-2$ is predicted to be  $a_t^{SM}=0.02$ \cite{Bernreuther:2005gq},  and the electric dipole moment is CP violating and is highly suppressed $d_t< 10^{-30}\ {\rm e}\cdot {\rm cm}$  \cite{Hoogeveen:1990cb}. 

Connecting those to the contributions from the electroweak dipole operators, we have the correspondence 
to the Wilson coefficients \cite{Haisch:2023upo}
\begin{equation}
a_V = \frac{2\sqrt{2}}{e}\frac{v m_t}{\Lambda^2} {\rm Re}[c_{t\gamma}] , \quad 
a_A = -\frac{2\sqrt{2}}{e}\frac{v m_t}{\Lambda^2} {\rm Im}[c_{t\gamma}] , 
\label{eq:ava}
\end{equation}
where the SM Higgs vacuum expectation value (vev) $v=0.246 ~\text{TeV}$. Notice $a_V$ is the real part of the Wilson coefficient, which is CP-even; and the $a_A$ is the imaginary part of the Wilson coefficient, which is CP-odd. 
%
When analyzing the Wilson coefficients, we only turn on one Wilson coefficient to be non-zero at a time for simplicity. 
We can thus have the following correspondence from Eq.~(\ref{eq:ava})

\begin{equation}
c_{t B} \approx  2.8\ a_V\ (\Lambda/\text{TeV})^2,\quad 
c_{t W} \approx 5.2\ a_V\ (\Lambda/\text{TeV})^2. 
\label{eq:av}
\end{equation}

\begin{table}[tb]
\centering
\begin{tabular}{|c||p{2.5cm}p{2.0cm}p{1.0cm}|p{2.5cm}c|c|c|c|}
\hline
&  \multicolumn{3}{c|}{$a_V ~~\to~~$ $\Lambda/ \sqrt{|c_{tB}|}$, $\Lambda/ \sqrt{|c_{tW}|}$ } & \multicolumn{2}{c|}{$|a_A|~~\to~~\Lambda/ \sqrt{|c_{tB}|}$} & Process & Ref. \\
&&[TeV] &[TeV]&&[TeV]&&\\
\hline
Current &\textcolor{black}
&0.91&2.9&-&-& LEP, LHC Run2 &\cite{Ellis:2020unq}\\
\hline
 & 6.75$\cdot 10^{-2}$&2.3 &1.7& 0.2792& 1.1& $\gamma g \to  t \bar t X$ (HL-LHC) & \cite{Koksal:2019gyo}\\
Future &$-2.34\cdot 10^{-2}$&3.9 &2.9& 
$2.77\cdot 10^{-2} $  &3.6&$\gamma 
e^- \to \bar t b \nu_e$ (CLIC)& \cite{Koksal:2016wlh} \\
 &2.0$\cdot 10^{-3}$&13&9.8&$ 2.06\cdot 10^{-2} $ &4.2&$\gamma\gamma \to t \bar t$ (CLIC)& \cite{Billur:2017yds} \\
 &2$\cdot 10^{-3}$&13&9.8 &- &-& $e^+ e^- \to t\bar t$ 
(ILC)& \cite{Amjad:2013tlv} \\
 &1.62$\cdot 10^{-3}$&15&11&$0.28  $&1.1& $e^+ e^- \to t\bar t$ (FCC-ee)& \cite{Janot:2015yza} \\
\hline
\end{tabular}
\caption{Lower bounds on the top-quark magnetic and electric dipole moments from $a_V$ and $a_A$ 
in Eq.~(\ref{eq:avaadef}) at $2\sigma$ from various current and future colliders (see the references therein), and the implication to our model-parameters $\Lambda/ \sqrt{|c_{t}|}$. 
}
\label{tab:constrain_av}
\end{table}

%

There are existing constraints at low energies on the top-quark electromagnetic dipole couplings from the vast literature.  
In Table \ref{tab:constrain_av}, we compile the current and future constraints from the existing results in terms of $a_V$ and $a_A$ from the hadron and lepton colliders. We relate the most stringent constraint of $a_V$ to the Wilson coefficient of our current interest $\Lambda/ \sqrt{c_{tB}}$ 
with Eq.~(\ref{eq:av}). 
We find that 
\begin{itemize}
\item the current LHC experiments on the $t\bar t X$ studies can already provide some interesting constraints with $\Lambda/ \sqrt{c_{t}} > 1$ TeV and 3 TeV for $c_{tB}$ and $c_{tW}$, respectively. The future HL-LHC could improve the sensitivity  with the process $\gamma g \to t\bar t$; 
\item  $e^+e^- \to t\bar t$ at the ILC and FCC-ee near the threshold production can be very sensitive to the dipole operator effects due to the enhanced interference with the chirality-flipping dipole operators, reaching $\Lambda/ \sqrt{c_{tB}} > 13$ TeV;
\item $\gamma\gamma \to t\bar t$ increases sensitivity at higher energies (CLIC), reaching some comparable results; 
\item bounds on $a_A$ are typically weaker because the contribution is from the imaginary part of the dipole operator.
\end{itemize}

Given a lower bound on the parameters $\Lambda/ \sqrt{c_{t}}$ 
denoted by $\lambda_c$ (in TeV), which is about 3 TeV from the LHC, it would translate to a relation 
\begin{equation}
c_{t B},\ c_{t W} <  ( {\Lambda / {\lambda_c\ \text{TeV}}})^2. 
\label{eq:relation}
\end{equation}
In practice in collider studies, the dipole operator cutoff scale must be above the collider energy for consistency of the EFT application. We thus assume 
\begin{equation}
\Lambda= 2 \sqrt{s} .
\label{eq:lambda}
\end{equation}
In anticipating potentially large new physics effects as motivated earlier, we will explore the possibly large Wilson coefficients satisfying the current bound $\lambda_c$ as
\begin{equation}
c_{t B},\ c_{t W} \approx {\rm min} 
[ ({{2 \sqrt s}/ {\lambda_c\ \text{TeV}}})^2, 4\pi] . 
\label{eq:c}
\end{equation}

\section{Sensitivity Reach at a Multi-TeV Lepton  Collider}
\label{sec:sen}

As discussed earlier, future colliders, such as the HL-LHC, ILC, CLIC, and FCC, would be able to improve the sensitivity to probe the top-quark dipole interactions. With the recent efforts of the feasibility studies of future colliders reaching a partonic center-of-mass energy of 10 TeV \cite{Narain:2022qud}, we hope to explore a new territory at the energy frontier. 
In this work, we study the sensitivity of exploring the electroweak dipole operators in Eq.~(\ref{eq:operator}) at a high energy muon collider \cite{Aime:2022flm,Black:2022cth}. 
For illustration, we will present the c.m.~energies 
\begin{equation}
\sqrt s= 1,\ 3\ {\rm and}\ 10\ 
{\rm TeV,\ with\ an\ integrated\  luminosity}\ 1,\  1,\ {\rm and}\ 
10\ {\rm ab}^{-1}, 
\label{eq:E}
\end{equation}
respectively.
 
\subsection{Production cross section for the signal channels}

There are many channels that can contribute to the measurements of the $t\bar t$ couplings to the gauge bosons $\gamma,Z, W$ via 
\begin{eqnarray}
\label{eq:tt}
&& \mu^+ \mu^- \to t\bar t,\\ 
\label{eq:tth}
&& \mu^+ \mu^- \to t\bar t h,\\ 
\label{eq:mmtth}
&& \mu^+ \mu^- \to \mu^+ \mu^- t\bar t,\ \nu  \bar\nu t\bar t.
\end{eqnarray}

The processes at lepton colliders, in particular the muon colliders, include the leading order top pair production in the $s$-channel in Eq.~(\ref{eq:tt}) as shown in Fig.~\ref{fig:Feynll_a}, which has the access to the full collider energy and thus potentially most sensitive to new physics. The total cross-section, however, generically falls at higher energies as $1/s$, and so does the leading interference between the SM and the dipole operators as $m_t^2/(s \Lambda^2)$.

The VBF contributions in Eq.~(\ref{eq:mmtth}) are via $\gamma/Z$ fusions in 
Fig.~\ref{fig:Feynll_b} and $WW$ fusion in Fig.~\ref{fig:Feynll_c}. Their production rates increase logarithmically over the colliding energy, and they thus hold the promise for exploring new physics at higher energies \cite{Han:2020uid,Liu:2023yrb,Barger:2023wbg}. The least explored channel perhaps is the $tth$ associate production in Eq.~(\ref{eq:tth}) as shown in Fig.~\ref{fig:Feynll_gh} and Fig.~\ref{fig:Feynll_fh}.
%
%
In particular, the 4-point coupling from the electroweak dipole operator in Fig.~\ref{fig:Feynll_fh} leads to the same chirality structure as that in the SM with highest energy growing behavior as $E^2/\Lambda^2$ and thus yields in a significant interference contribution. One would expect to have good sensitivity to the BSM effects at high energies. 

Our calculations are performed based on the UFO model file \cite{Brivio:2017btx} generated by FeynRules \cite{Alloul:2013bka} in the leading order. We apply the model file in MadGraph5 \cite{Alwall:2014hca} with the default SM parameter choices. We use Whizard \cite{Kilian:2007gr} for the processes involving muon beam initial state radiation (ISR). 
%
%
We formally express the cross-section including the SM expectations and the BSM new physics (NP) contributions. 
%
  \begin{eqnarray}
\sigma_{\text{total}}&=& \text{SM}^2 + 2 | \text{SM} * \text{NP}|+ \text{NP}^2  \nonumber
\\
&\equiv &\sigma_{SM}+\sigma_{NP}^{(1)}+\sigma_{NP}^{(2)}\  . 
\label{eq:NP}
\end{eqnarray}
In general the interference term
is the leading effect for BSM, and the NP square term is power-suppressed at an energy $E$ by $E^2/\Lambda^2$. If it is not parametrically smaller, then the higher order operators, especially at dim-8 in Eq.~(\ref{eq:SMEFT}), should be included.

\begin{figure}[tb]
    \centering
    \subfigure[]{\includegraphics[width=0.25\textwidth]{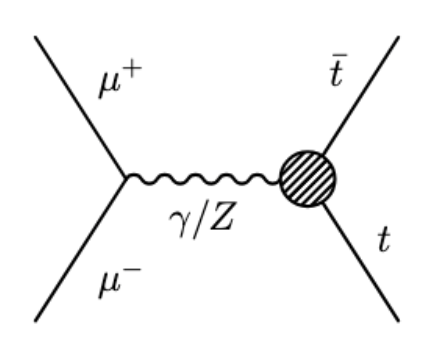}
    \label{fig:Feynll_a}
    }
    \subfigure[]{\includegraphics[width=0.25\textwidth]{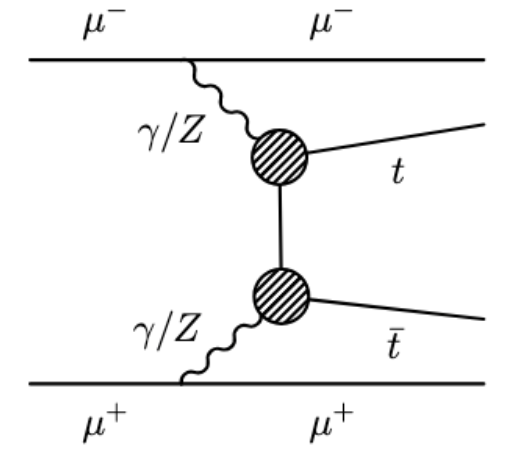}
    \label{fig:Feynll_b}
    }
    \subfigure[]{\includegraphics[width=0.25\textwidth]{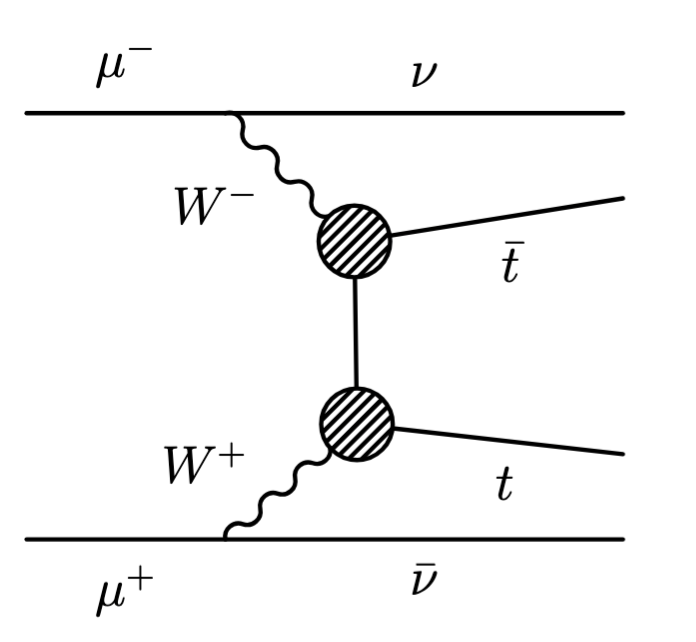}
    \label{fig:Feynll_c}
    } 
       \subfigure[]{\includegraphics[width=0.26\textwidth]{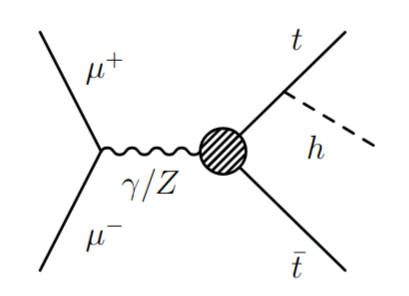}
     \label{fig:Feynll_gh}
    }
    \subfigure[]{\includegraphics[width=0.26\textwidth]{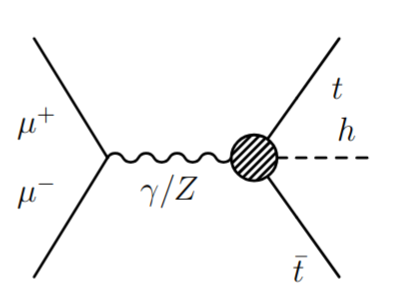}
     \label{fig:Feynll_fh}
   }    
    \caption{Representative Feynman diagrams for $t \bar t$ and $t\bar t H$ production in $\mu^+\mu^-$ collisions. The circles denote the possible new physics contribution. 
    }
    \label{fig:Feynll}
\end{figure}

We first show in Fig.~\ref{fig:mumu_Ecm} the total cross-section involving $t\bar t$ production at a muon collider as in Eqs.~(\ref{eq:tt})$-$(\ref{eq:mmtth}) 
versus the c.m.~energy $\sqrt s$ from 1 TeV to 10 TeV. The dashed curves are the SM expectation and the solid lines include the dipole operators up to terms of $1/ \Lambda^4$, including the effects from the initial state radiation (ISR). The ISR effects turn out to be quite significant in our treatment since the effective collision energy would be reduced, leading to about 
10\%$-$20\% increase of the $s$-channel production cross section and some decrease of the energy-dependent NP signal. For illustration in the calculation, we have chosen the dipole operator cutoff scale above the collider energy as in Eq.~(\ref{eq:lambda}) 
%
%
with $c_{t B}=1$. 
Not far above the threshold, the $s$-channel processes $\mu^+ \mu^- \to t \bar t$ and $\mu^+ \mu^- \to t \bar t h$ dominate the cross section, and then fall as $1/s$. 
We see in Fig.~\ref{fig:mumu_Ecm} that the effects due to the dipole operators in the $t \bar t$ processes become smaller at higher energies, and are significantly weaker than those in $t \bar t h$ processes. The reason for those can be understood from the fact that the leading dipole operator effect comes from the interference as illustrated in Eq.~(\ref{eq:NP}). Because of the chirality flip of the dipole operator, the interference with the SM $t\bar t$ is suppressed by the mass insertion as $m_t/E$. In contrast, the interference with the SM $t\bar th$, which already has a chirality flip due to the Yukawa coupling, does not suffer from this suppression. We summarize the scaling behavior in Table \ref{tab:Modelmuon_xsec}. Some related analyses for the scaling behavior can be found in Ref.~\cite{Chen:2022yiu}. 

The vector boson fusion processes (VBF)  
exhibit the log$(s)$ increasing behavior as expected. They take over the $t\bar t$ production near $\sqrt s \approx 4-5$ TeV, as seen in Fig.~\ref{fig:mumu_Ecm}. To assure the VBF contributions in the $\mu^+\mu^- t\bar t, \ \nu\nu t \bar t$ channels and remove the contamination from $Z\to \mu^+\mu^-,\ \nu\bar\nu$, we impose a selection cut for the rest of the analyses 
\begin{equation}
M_{\ell\ell},\ M_{\nu\nu}> 150\ {\rm GeV.}
\label{eq:cuts}
\end{equation}
The neutral current VBF from $\gamma/Z$ has a larger cross-section than the charged current VBF. This can be traced back to the co-linear enhancement for a photon radiation as $\log(Q^2/m_\mu^2)$ versus a massive gauge boson radiation as $\log(Q^2/M_W^2)$. Consequently, within the energy range of our current interest, we will ignore the dipole contributions in the process $\mu^+\mu^- \to  \nu \bar\nu t \bar t$ for the signal analysis. 

\begin{figure}[t]
    \centering
   \includegraphics[width=0.75\textwidth]{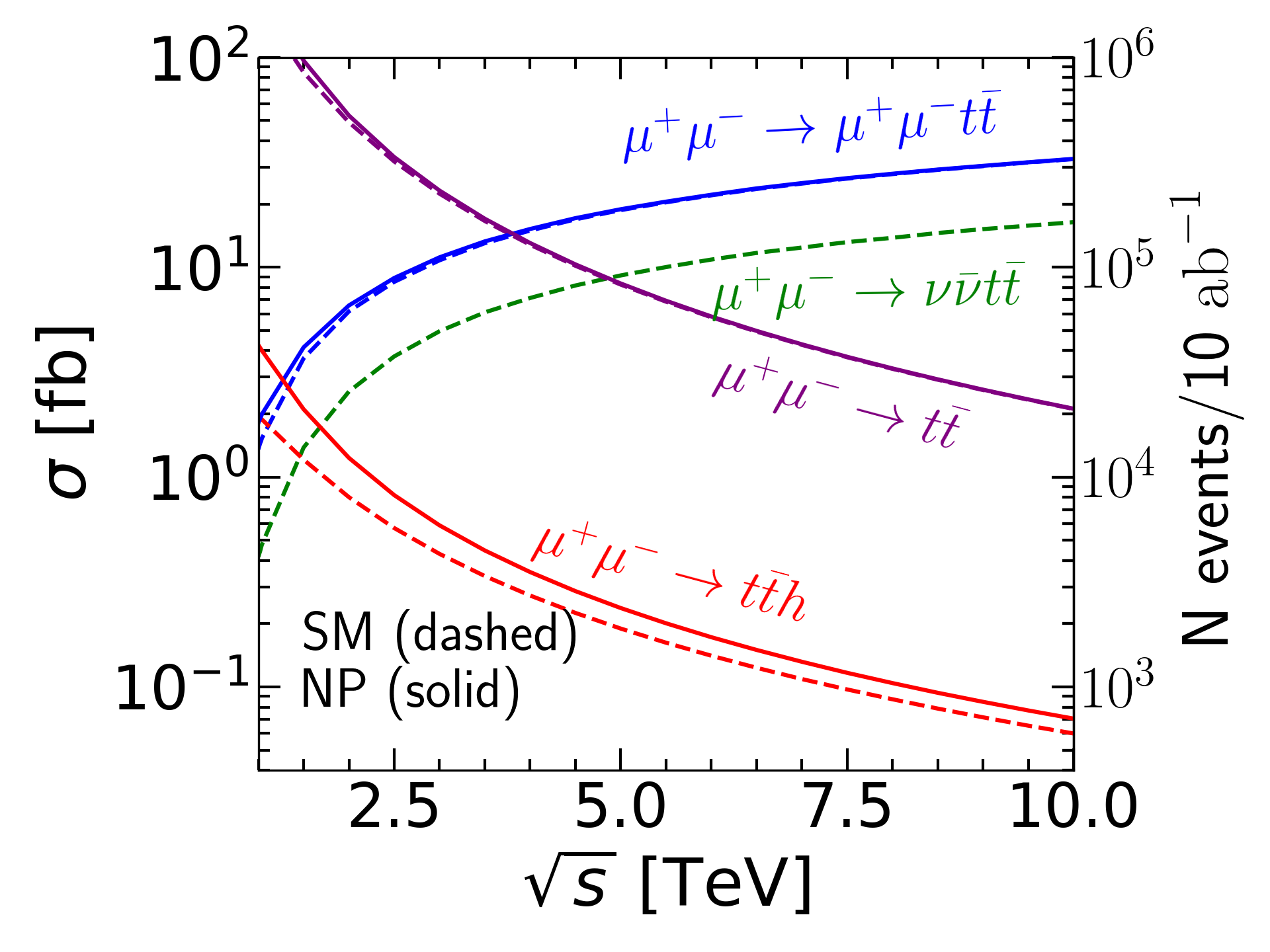}
    \caption{
    The total cross-section for $\mu^+\mu^- \to t\bar t+X$ production versus the c.m.~energy $\sqrt s$ including the ISR effects. The dashed curves are the SM expectation and the solid lines include the dipole operators keeping terms of $1/ \Lambda^4$, for $c_{t B}=1$, $\Lambda= 2 \sqrt{s}$.
    }
\label{fig:mumu_Ecm}
\end{figure}

\begin{table}[tb]
\centering   
\begin{tabular}{|>{\centering\arraybackslash}p{4cm}|>{\centering\arraybackslash}p{3cm}|>{\centering\arraybackslash}p{3cm}|>{\centering\arraybackslash}p{3cm}|}
\hline
Processes &$\sigma_{SM}$&  $\sigma_{NP}^{(1)}$ &$\sigma_{NP}^{(2)}$ \\
\hline
$\mu^+ \mu^- \to  t \bar t $ & $\frac{1}{E^2} $& $\frac{1}{E^2} \frac{m_t^2}{\Lambda^2}$ & $\frac{1}{E^2} \frac{m_t^2 E^2}{\Lambda^4}$ \\
\hline
$\mu^+ \mu^- \to  t \bar t h$ & $\frac{1}{E^2} $& $\frac{1}{E^2} \frac{E^2}{\Lambda^2}$ & $\frac{1}{E^2} \frac{E^4}{\Lambda^4}$ \\
\hline
\end{tabular}
\caption{Energy scaling behavior at the central region. 
}
\label{tab:Modelmuon_xsec}
\end{table}

%

\begin{figure}[tb]
\centering
\includegraphics[width=0.49\textwidth]{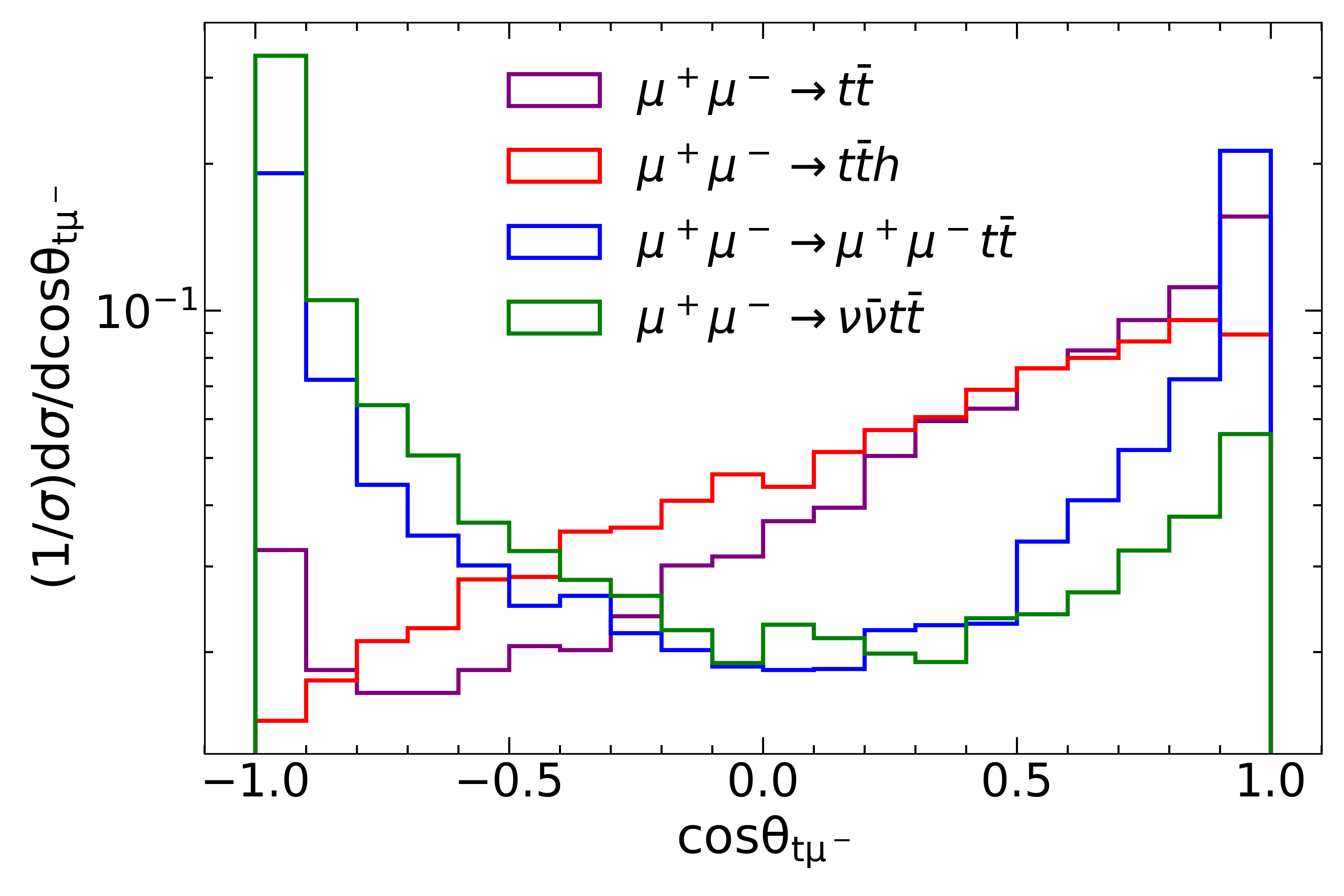}
\includegraphics[width=0.49\textwidth]{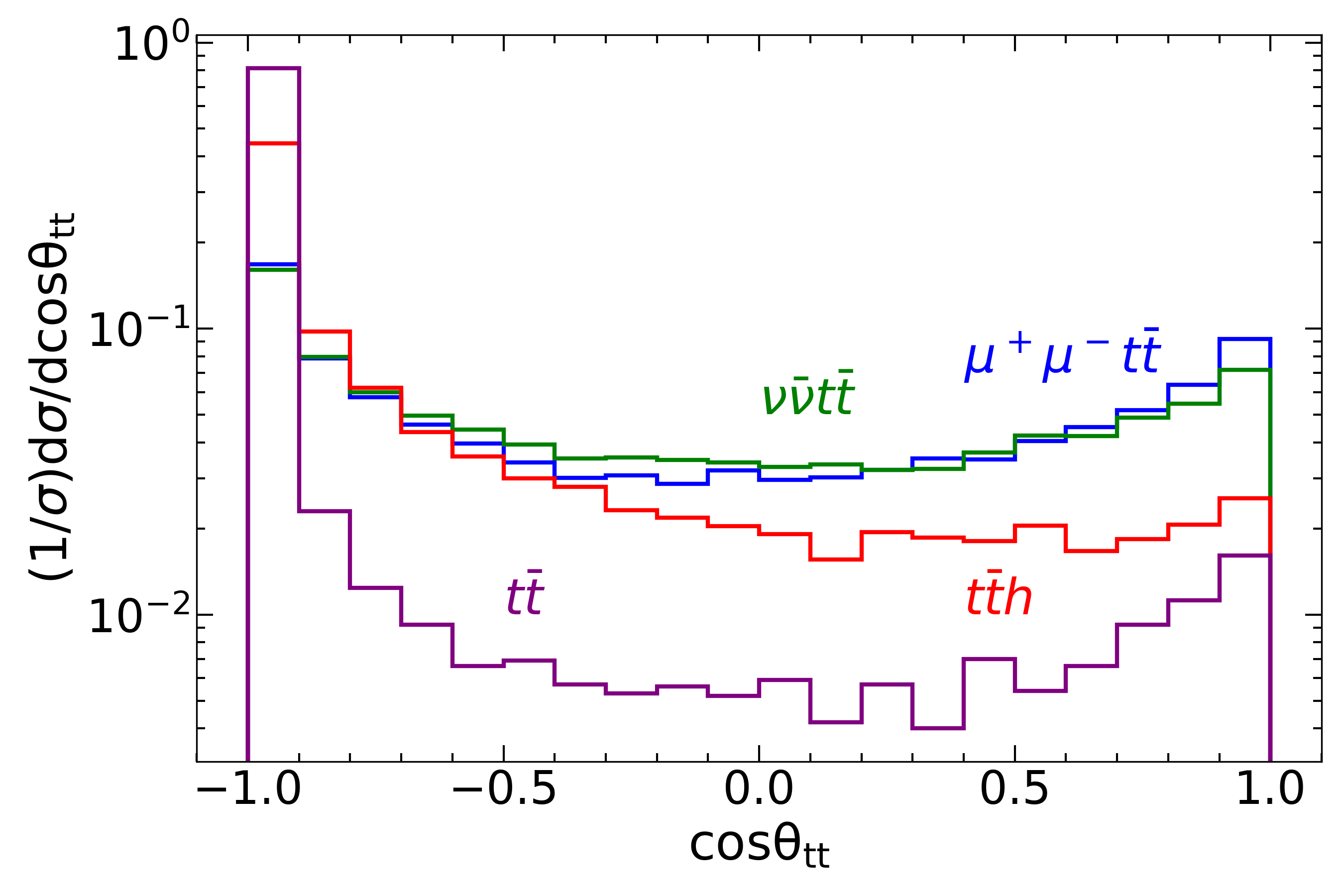}
\includegraphics[width=0.49\textwidth]{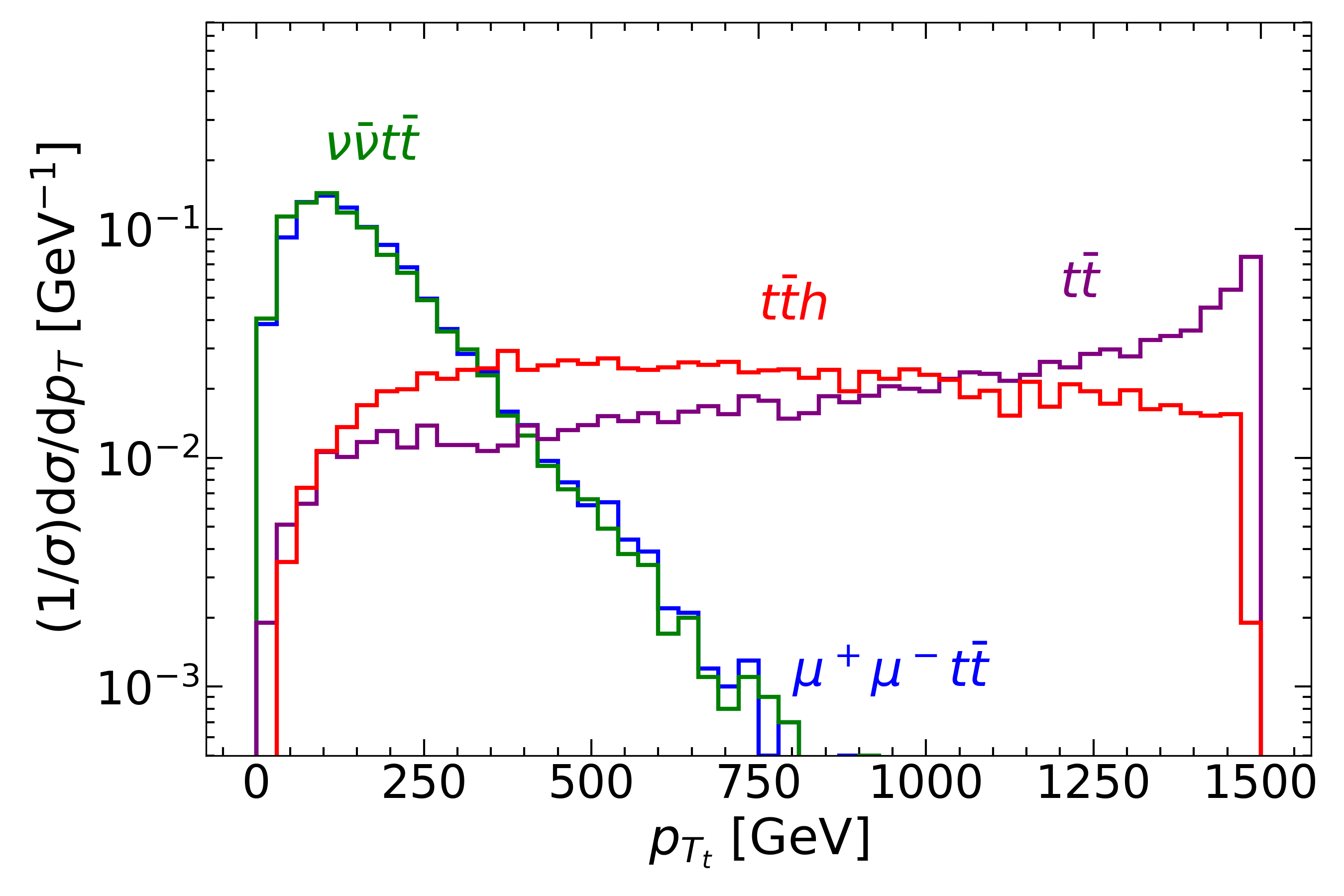}
\includegraphics[width=0.49\textwidth]{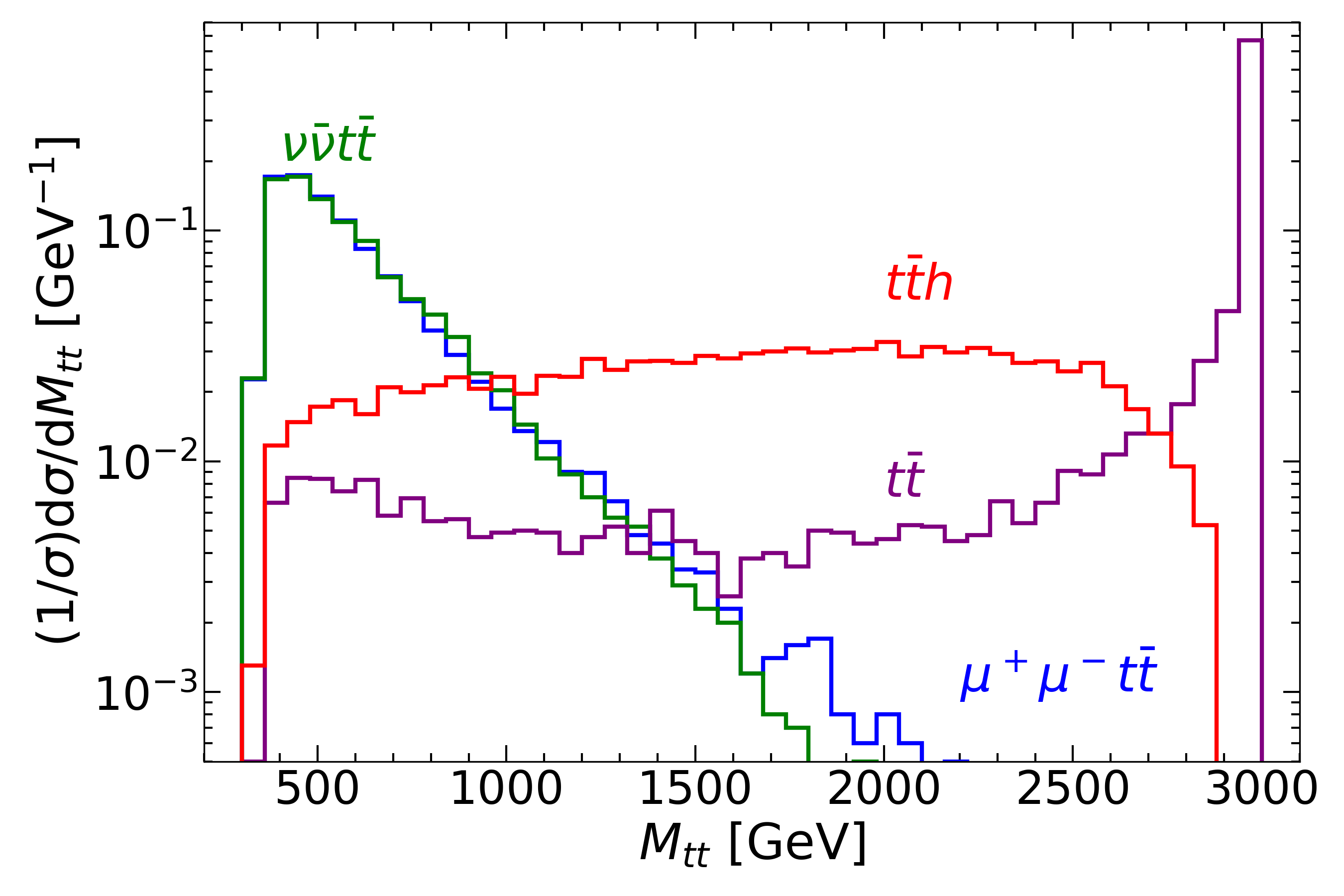}
\caption{Normalized kinematic distributions for the opening angles of 
$\cos\theta_{t\mu^-}$ and 
$\cos\theta_{tt}$, the
transverse momentum $p_T^{t}$ and the invariant mass $M_{tt}$ at $\sqrt s=3$ TeV. 
}
\label{fig:dist_mumu2tt}
\end{figure}

It is instructive to show some kinematical features of the processes.  In Fig.~\ref{fig:dist_mumu2tt}, we present the normalized distributions of (a) the angle of the top with respect to the 
$\mu^-$ beam cos$\theta_{t\mu^-}$, (b) the opening angle between $t$ and $\bar t$ cos$\theta_{tt}$, (c) the transverse momentum of top quark $p_T(t)$, and the invariant mass of the $t\bar t$ system $M_{tt}$, at a 3 TeV muon collider. The inclusion of the ISR is the cause for the long tail of the invariant mass peak for the $M_{tt}$ and the Jacobean peak for $p_T(t)$, respectively. For the VBF processes, cuts to remove the $Z\to \ell\ell$ resonant contributions as in Eq.~(\ref{eq:cuts}) have been applied. 
For moderate size of new dipole contributions, they are not appreciable on those normalized distributions.
%
The asymmetry of the top quark forward angle distribution shows the parity violation induced by the $Z$ boson mediator. 
The opening angle distribution between the top pair is dominantly back to back.

\subsection{$\mu^+\mu^-\to t\bar t$ channel}
For the $s$-channel $\mu^+ \mu^- \to t \bar t$ production, 
as discussed earlier, 
the leading new physics interference term $\sigma_{NP}^{(1)}$ is helicity suppressed as 
$m_t/E \rightarrow 0$. The NP squared term $\sigma_{NP}^{(2)}$ is almost constant with changing collision energy. We show some representative cross-sections in Table \ref{tab:Modelmu_tree_xsec10}, 
for three choices of collider energies $\sqrt s=1$ TeV, 3 TeV, and 10 TeV. We wish to present results with the larger values of the Wilson coefficients allowed, and a sensible scale as guided by Eqs.~(\ref{eq:lambda}) and (\ref{eq:c}). For the sake of numerical illustration, we have chosen $(c_{tB}, \Lambda) = (0.45,\ 2\ {\rm TeV}),\ (4,\ 6\ {\rm TeV})\ {\rm and}\ (4\pi,\ 20\ {\rm TeV})$, respectively. We have not shown the results for the contributions from $c_{tW}$ since the signal is slightly weaker due to the smaller $\gamma\gamma$ initial state contribution from $c_{t\gamma}$ in Eq.~(\ref{eq:mix}), and the general behavior is similar. 
As expected, we observe the decreasing of the SM cross section and the interference contribution $\sigma_{NP}^{(1)}$ as $1/s$. For a rough guide, the signal-to-background ratios are also shown in the last row of the table. 

\begin{table}[tb]
\centering
\begin{tabular}{|c||c|c|c|c|c|c|c|c|c|}
\hline
$\sigma$[pb]&\multicolumn{3}{c|}{ $\mu^+ \mu^- \to  t \bar t $ }& \multicolumn{3}{c|}{ $\mu^+ \mu^- \to  t \bar t h $ }& \multicolumn{3}{c|}{ $\mu^+ \mu^- \to  \mu^+ \mu^-  t \bar t $ }\\
\hline
$\sqrt{s}$&1 TeV & 3 TeV& 10 TeV&1 TeV & 3 TeV& 10 TeV&1 TeV & 3 TeV& 10 TeV\\
$\Lambda$& 2 TeV & 6 TeV& 20 TeV&2 TeV & 6 TeV& 20 TeV& 2 TeV & 6 TeV& 20 TeV  \\
$c_{tB}$ & 0.45& 4 & $4\pi$ &0.45 & 4 & $4\pi$ & 0.45& 4 & $4\pi$\\
\hline\hline
$\sigma_{SM}$ 
&0.18&2.2 $\cdot 10^{-2}$&2.1$\cdot 10^{-3}$
&1.9$\cdot 10^{-3}$&4.3$\cdot 10^{-4}$&6.0$\cdot 10^{-5}$
 &1.4$\cdot 10^{-3}$& 1.1$\cdot 10^{-2}$&2.4$\cdot 10^{-2}$
\\
$\sigma_{\textrm{NP}}$
&0.02&0.5 $\cdot 10^{-2}$&0.2$\cdot 10^{-3}$
&0.9$\cdot 10^{-3}$&1.1$\cdot 10^{-3}$&4.7$\cdot 10^{-4}$
&0.2$\cdot 10^{-3}$& 0.1$\cdot 10^{-2}$&0.1$\cdot 10^{-2}$ \\
$\sigma_{\textrm{total}}$
&0.20&2.7 $\cdot 10^{-2}$&2.3$\cdot 10^{-3}$
&2.8$\cdot 10^{-3}$&1.5$\cdot 10^{-3}$&5.3$\cdot 10^{-4}$
&1.6$\cdot 10^{-3}$& 1.2$\cdot 10^{-2}$&2.5$\cdot 10^{-2}$
\\
\hline
S/B
&0.11&0.23&0.095 
& 0.47&2.6&7.8
&0.14&0.091&0.042
\\
\hline
\end{tabular}
\caption{
Representative cross sections for the SM background and the dipole operator signal with a variety of choices of the Wilson coefficients $c_{tB}$ and the scale $\Lambda$, with the collider energies 1 TeV, 3 TeV, and 10 TeV. 
The ISR effects are included and acceptance cuts in Eq.~(\ref{eqn:cuts}) are applied. 
}
\label{tab:Modelmu_tree_xsec10}
\end{table}


Besides the leading SM background $\mu^+\mu^- \to t \bar t$, there are some other potential backgrounds, as shown in Fig.~\ref{fig:Feynllbg}. 
We tabulate their production cross sections at a 3 TeV collider in Table \ref{tab:bgxsec}.
Although the production rates for those background processes are comparable to the leading process, they are kinematically quite different from the signal at hand. For instance, if we demand the top quark reconstruction $m(Wb)\approx m_t$, the background $\mu^+ \mu^- \to t b W $ shown 
in Fig.~\ref{fig:Feynll_f} will be highly suppressed. Similarly, if we require the $t \bar t$ mass reconstruction and the $b \bar b$ is not from a $Z$ decay at $M_Z$, the background $\mu^+ \mu^- \to W W Z $ in Fig.~\ref{fig:Feynll_g} can be effectively removed. Therefore, we will not consider these background in the following numerical analyses. 

\begin{figure}[tb]
    \centering
    \subfigure[]{\includegraphics[width=0.34\textwidth]{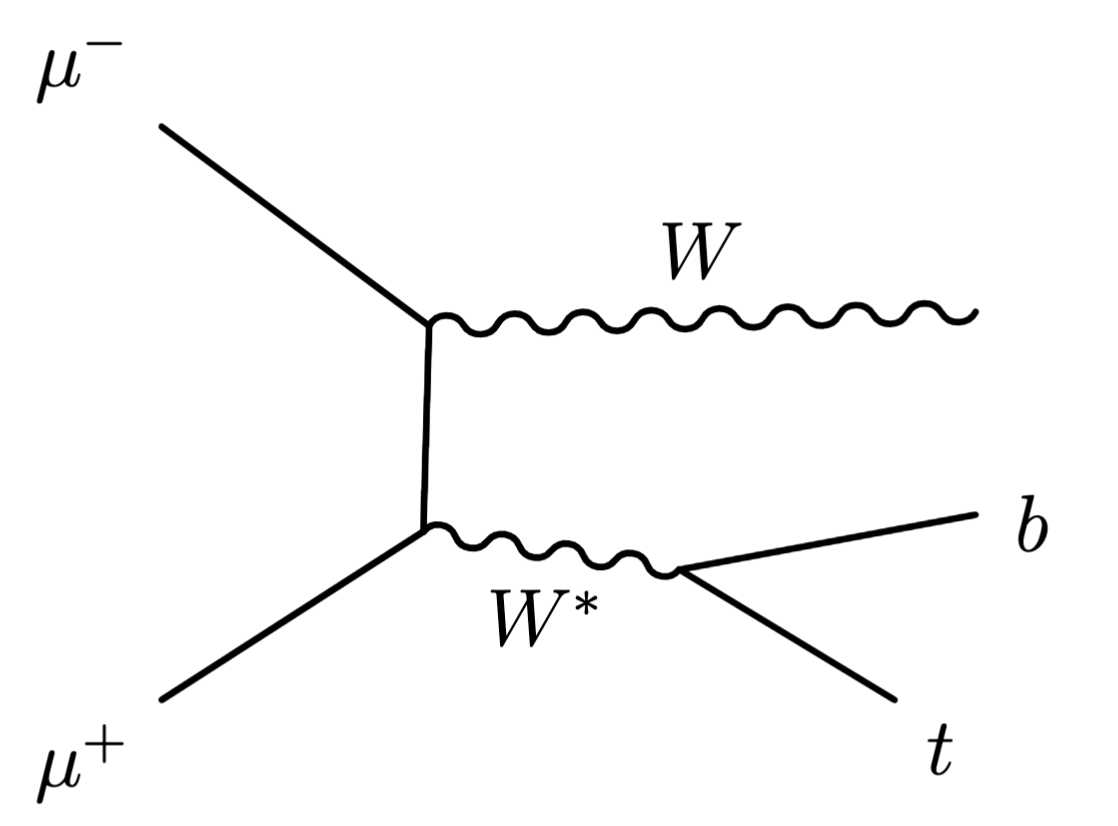}
    \label{fig:Feynll_f}
    }
    \subfigure[]{\includegraphics[width=0.34\textwidth]{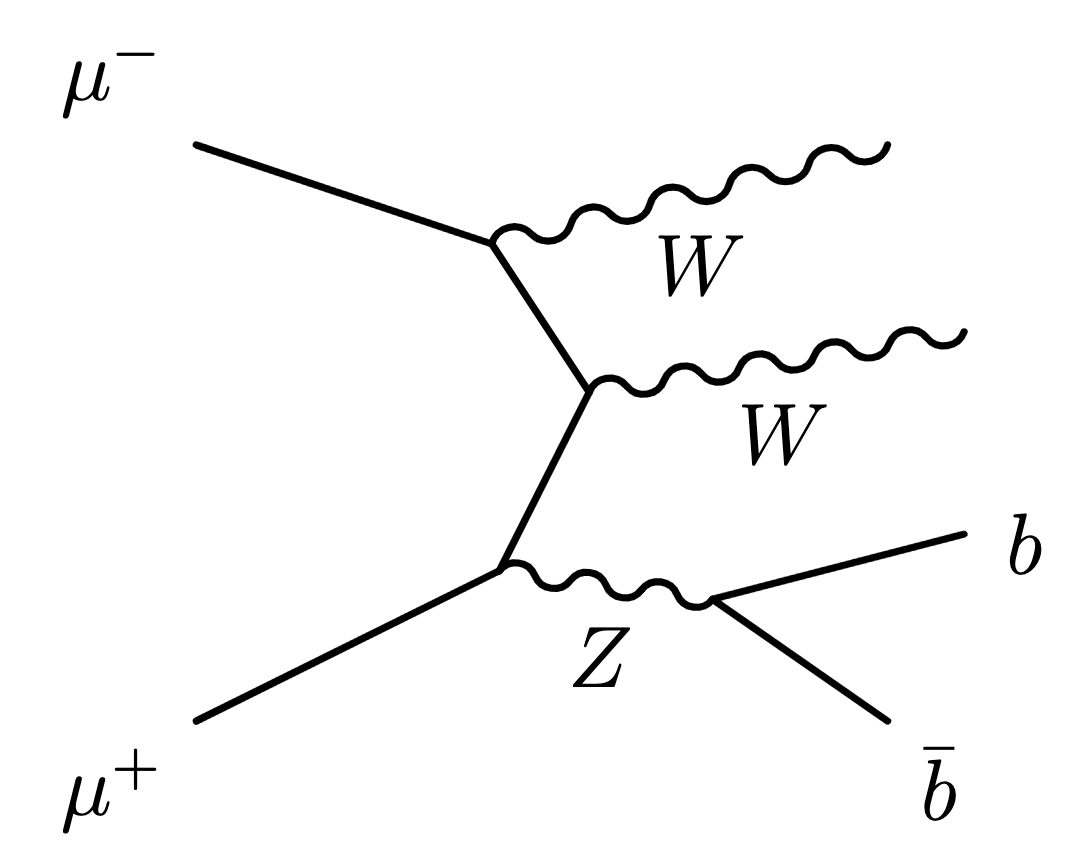}
     \label{fig:Feynll_g}
    }
    \caption{Background Feynman diagrams for the $t\bar t$ final state at a muon collider.}
    \label{fig:Feynllbg}
\end{figure}

\begin{table}[tb]
\centering
\begin{tabular}{|c|c|}
\hline
Processes& SM cross-section [pb]\\
\hline
$\mu^+ \mu^- \to  t \bar t $ & 2.0$\cdot 10^{-2}$\\
\hline
$\mu^+ \mu^- \to  \bar t b W^+ $ &8.8$\cdot 10^{-3}$\\
$\mu^+ \mu^- \to  W^+ W^- Z $ & 3.1$\cdot 10^{-2}$\\
\hline
\end{tabular}
\caption{Background cross sections at  $\sqrt{s}=3$ TeV without kinematic cuts. 
}  
\label{tab:bgxsec}
\end{table}



\subsection{$\mu^+\mu^-\to t\bar t h$ channel}

As discussed earlier, the new physics effect of the dipole operators decreases versus energy for the process $\mu^+\mu^- \to t\bar t$ due to the diminishing interference. In contrast, however, the NP effect from the process $\mu^+ \mu^- \to t \bar t h$ would be enhanced at higher energies, as shown in the diagrams of Figs.\ref{fig:Feynll_gh} and \ref{fig:Feynll_fh}. 
Again, we show some representative cross-sections in Table \ref{tab:Modelmu_tree_xsec10}, 
for three choices of collider energies $\sqrt s=1$ TeV, 3 TeV, and 10 TeV and a few values of the Wilson coefficients and the scale. As expected from the analysis of Table \ref{tab:Modelmuon_xsec}, the interference term $\sigma_{NP}^{(1)}$ is almost constant with increasing collision energy, and the NP square term $\sigma_{NP}^{(2)}$ increases with the collision energy. 
One would therefore expect stronger sensitivity for the dipole operator effects at a higher energy collider, as already evident by $S/B$. In particular, exploring the electroweak dipole effects at lepton colliders possesses a unique opportunity since the same effects would be difficult to stand out of the large QCD backgrounds in hadronic collisions. 


\subsection{$\mu^+\mu^-\to \mu^+\mu^- t\bar t$ channel}

At higher energies, the vector boson fusion processes are logarithmically enhanced and the VBF contributions in Eq.~(\ref{eq:mmtth}) take over, as seen in Fig.~\ref{fig:mumu_Ecm}.
%
%
As discussed earlier, the neutral current VBF from $\gamma/Z$ has a larger cross-section than the charged current VBF by more than a factor of 2. We find that the effect of $c_{tW}$ on the charged current process is also weaker than that on the neutral current. Therefore, we will ignore the dipole contributions in the process $\mu^+\mu^- \to  \nu \bar\nu t \bar t$ for the signal analysis. 
On the other hand, since the outgoing $\mu^+\mu^-$ are extremely forward at a multi-TeV collider, we would not demand the $\mu^\pm$ tagging in the forward region. As such, we will have to include the contributions of $\mu^+\mu^- \to  \nu \bar\nu t \bar t$ in our background analysis. 
The signal and background cross sections are summarized in Table \ref{tab:Modelmu_tree_xsec10}, for selective collider energies and model parameter choices. We see the prospects for this channel to probe the dipole operators are comparable to the other two. We note, however, that this channel is not as competitive as that at CLIC   \cite{Billur:2017yds}, due to the much larger $\gamma\gamma$ flux in $e^+e^-$ collisions. 

\subsection{Sensitivity} 

Based on the previous discussions for the signal from the electroweak dipole operators and the background expectation, we set to calculate the sensitivity for the above processes. 
We estimate the events for the signal $(S)$ and background $(B)$ as
\begin{align}
\label{eq:S}
S &= N_{\textrm{total}} - N_{\textrm{SM}} = \mathcal{L} \cdot (\sigma_{\textrm{total}} - \sigma_{\textrm{SM}}) \cdot K = \mathcal{L} \cdot \sigma_{\textrm{NP}} \cdot K, \\ 
\label{eq:B}
B &= N_{\textrm{SM}} = \mathcal{L} \cdot \sigma_{\textrm{SM}} \cdot K.
\end{align}

where $K$ is the efficiency factor including the top decay branching fractions and experimental acceptance efficiency. For the sake of effective event reconstruction, we consider the hadronic decays of the top quark ($bjj, \bar bjj$) and the Higgs boson ($b\bar b$). 
The branching fractions and tagging efficiency are as follows \cite{CMS-DP-2023-065} 
\begin{eqnarray}
&& \text{BR}(t \to W b)=0.95,\  \text{BR}(W \to j j)=0.68,\  \text{BR} (H \to b b)=0.53 , \\
&& \text{top-tag}=0.81,\  \text{Higgs-tag}=0.75 .
\end{eqnarray}
We also impose the rapidity acceptance cuts to simulate the detector coverage as $10^\circ$ with respect to the beam axis 
\begin{eqnarray}
|\eta_t|<2.44,\quad  |\eta_h|<2.44 .
\label{eqn:cuts}
\end{eqnarray}
%
In the clean lepton collider environment, we ignore the systematic uncertainties, that can be controlled to a sub-percentage level \cite{Accettura:2023ked}. 
We thus estimate the signal observability by the statistical standard deviation of Eqs.~(\ref{eq:S}) and (\ref{eq:B}) 
\begin{eqnarray}
\chi = \frac{S}{\sqrt{B}}
\label{eq:chi}
\end{eqnarray}
In the following, we adopt $\chi=2$ for $2 \sigma$ sensitivity limit, which is equivalent to a 95\% Confidence Level (CL).

In extracting the new physics effects from the dipole operators, we follow the discussions in Eq.~(\ref{eq:NP}), and express the new physics contributions in terms of an interference and a squared term 
\begin{equation}
{\sigma_{NP} \over \sigma_{SM}} =  {A \over \Lambda^2} + {B \over \Lambda^4}\ . 
\label{eq:AB}
\end{equation}
where $\Lambda$ is in a unit of TeV. 
For a given collider energy, we calculate the cross sections with two different $\Lambda$ values and then fit $A$ and $B$ for new physics contributions.

Our results are summarized in Table \ref{tab:Modelmuon_sum} for representative choices of the c.m.~energies as given in Eq.~(\ref{eq:E}), as well as two threshold energies for comparison $\sqrt s = 365$ GeV for $t\bar t$ and 500 GeV for $t\bar t h$. For the signal evaluation, we turn on the Wilson coefficients $c_{tB}$ or $c_{tW}$ to be non-zero one at a time. We extract the new physics contributions in the form of $A$ and $B$ as in Eq.~(\ref{eq:AB}). We then fit the $2 \sigma$ sensitivity $\chi$ in Eq.~(\ref{eq:chi}), including the ISR effects and the efficiency factor $K$ to count for the branching fractions and reconstruction efficiency mentioned above.

We observe several important features from our results. First of all, closely following the scaling behavior as in Table  \ref{tab:Modelmuon_xsec}, the leading interference term ($A/\Lambda^2$) in $t \bar t\ (\mu^+\mu^- t\bar t$) channel remains essentially constant with respect to the SM cross section over energy, while the quadratic term ($B/\Lambda^4$) goes like $E^2$. In contrast to the $t \bar t h$ channel, both terms go like $E^2$ and $E^4$, respectively. Realizing the helicity suppression in the interference term, the $t \bar t$ channel is more sensitive to the dipole operator effect near the threshold, while the $t \bar t h$ channel is more sensitive at higher energies. The $\mu^+\mu^- t\bar t$ channel contributes moderately, and the $\nu \bar\nu t\bar t$ channel is to an even lesser extent. As shown in the last two columns in Table \ref{tab:Modelmuon_sum}, we may be able to approach a new physics scale above the collider energy for $c_{tB},\ c_{tW} \approx 1$. We note that one reaches a slightly better sensitivity for $c_{tB}$ than that for $c_{tW}$. This can be traced down to the larger $\gamma\gamma$ contribution due to the weak mixing effects as in Eq.~(\ref{eq:mix}) from $c_{t\gamma}$. 
We would like to emphasize the merit of the lepton colliders to probe the electroweak operators, since 
at hadron colliders such as the LHC, we are unable to reach the similar sensitivity shown here because the electroweak effects will be overshadowed by the QCD processes, with additional large SM backgrounds.

Thus far, we have assumed that the Wilson coefficients $c_{tB},\ c_{tW}$ are both real, which correspond to the magnetic dipole effects at low energies as seen in Eq.~(\ref{eq:ava}). It is of special interest to ask if one can also probe the electric dipole effects when these operators contain a complex phase. We carry out the similar analysis for the signal and background as before and take $c_{tB}$ to be an imaginary value. As such, there is no interference between the SM and a dipole operator with a complex phase, and the leading contribution is at the order of $B/\Lambda^4$. Our results are summarized in Table \ref{tab:Modelmuon2}. We see that the sensitivity is not as strong as the magnetic dipoles, especially at low energies when the leading interference term is much more dominant but absent here. 

\begin{table}[tb]
\centering   
\noindent\hskip-0.5cm
\begin{tabular}{c||c|c|p{1.5cm}c|cc|c|c}
\hline
 &$\sqrt{s}$&$\sigma_{SM}$& \multicolumn{4}{c|}{ $\sigma_{NP}/\sigma_{SM}= A/\Lambda^2 + B/ \Lambda^4 $ } &\multicolumn{2}{c}{ $2 \sigma$ value [TeV]}\\
\hline
Processes& TeV &[pb]& $A_{tB}$ & $B_{tB}$ &  $A_{tW}$ & $B_{tW}$ & $\Lambda/\sqrt{|c_{tB}|}$ & $\Lambda/\sqrt{|c_{tW}|}$\\
\hline\hline
 &0.365&0.44&0.87&0.27&0.72&0.19&12&11\\
 $\mu^+ \mu^- \to  t \bar t $ &1&0.18&1.0&0.97&0.79&0.64&11&9.4\\
with ISR&3&2.2$\cdot 10^{-2}$&1.0&6.8&0.86&4.6&6.8&6.2\\
~~~~~~~&10&2.1$\cdot 10^{-3}$&1.1&71&0.82&47&8.7&7.8\\
\hline
 &0.5&2.4$\cdot 10^{-4}$&1.7&1.0&1.3&0.70&2.2&2.0\\
$\mu^+ \mu^- \to  t \bar t h $ &1&1.9$\cdot 10^{-3}$&3.4&5.2&2.7&3.5&5.1&4.6\\
with ISR &3&4.3$\cdot 10^{-4}$&10&107&8.0&70&6.6&5.9\\
 &10&6.0$\cdot 10^{-5}$&57&6.2$\cdot 10^{3}$&38&4.1$\cdot 10^{3}$&18&15\\
\hline
$\mu^+ \mu^- \to \mu^+ \mu^- t \bar t $ + &1&1.8$\cdot 10^{-3}$&1.1&0.90&0.63&0.30&3.2&2.4\\
 $\mu^+ \mu^- \to \nu \bar\nu t \bar t $ (BG)&3&1.6$\cdot 10^{-2}$&1.0&1.3&0.60&0.46&4.9&3.7\\
&10&3.7$\cdot 10^{-2}$&1.1&3.0&0.57&1.1&11&7.9\\
\hline
\end{tabular}
\caption{Summary results for the $2\sigma$ sensitivity on $\Lambda/\sqrt{|c_{tB}|}$ and $\Lambda/\sqrt{|c_{tW}|}$ for real $c_{tB}$ and $c_{tW}$. The collider c.m.~energies and luminosities are as in Eq.~(\ref{eq:E}). The results near thresholds are also presented for $t\bar t$ at $\sqrt s =365$ GeV and for $t\bar t h$ at 500 GeV, both with 1 ab$^{-1}$ luminosity. The ISR effects are included and all acceptance cuts discussed in the text are applied. 
}
\label{tab:Modelmuon_sum}
\end{table}

\begin{table}[tb]
\centering   
\begin{tabular}{c||c|c|c| c}
\hline
 &$\sqrt{s}$&$\sigma_{SM}$& \multicolumn{1}{c|}{ $\sigma_{NP}/\sigma_{SM}= B/ \Lambda^4 $ } & $\Lambda/\sqrt{|c_{tB}|}$ value at $2 \sigma$\\
\hline
Processes& TeV &[pb]&  B &  TeV\\
\hline\hline
&0.365&0.44&0.01&1.2\\
  &1&0.18&0.67&2.9\\
$\mu^+ \mu^- \to  t \bar t $ &3&2.2$\cdot 10^{-2}$&6.7&4.0 \\
with ISR&10&2.1$\cdot 10^{-3}$&71&7.2\\
\hline
&0.5&2.4$\cdot 10^{-4}$ &0.02&0.47\\
 $\mu^+ \mu^- \to  t \bar t h $&1&1.9$\cdot 10^{-3}$&1.8&1.9\\
with ISR  &3&4.3$\cdot 10^{-4}$&87&4.1 \\
&10&6.0$\cdot 10^{-5}$&6.1$\cdot 10^{3}$&13\\
\hline
$\mu^+ \mu^- \to \mu^+ \mu^- t \bar t $ +  &1&1.8$\cdot 10^{-3}$&0.66&1.5\\
$\mu^+ \mu^- \to \nu \bar\nu t \bar t $ (BG)&3&1.6$\cdot 10^{-2}$&1.1&2.2\\
&10&3.7$\cdot 10^{-2}$&2.8&4.1\\
\hline
\end{tabular}
\caption{Summary results for the $2\sigma$ sensitivity on $\Lambda/\sqrt{|c_{tB}|}$ for imaginary $c_{tB}$. The collider c.m.~energies and luminosities are as in Eq.~(\ref{eq:E}). The results near thresholds are also presented for $t\bar t$ at $\sqrt s =365$ GeV and for $t\bar t h$ at 500 GeV, both with 1 ab$^{-1}$ luminosity. The ISR effects are included and all acceptance cuts discussed in the text are applied. 
}
\label{tab:Modelmuon2}
\end{table}

\begin{figure} [tb]
    \centering
     \includegraphics[width=0.75\textwidth]{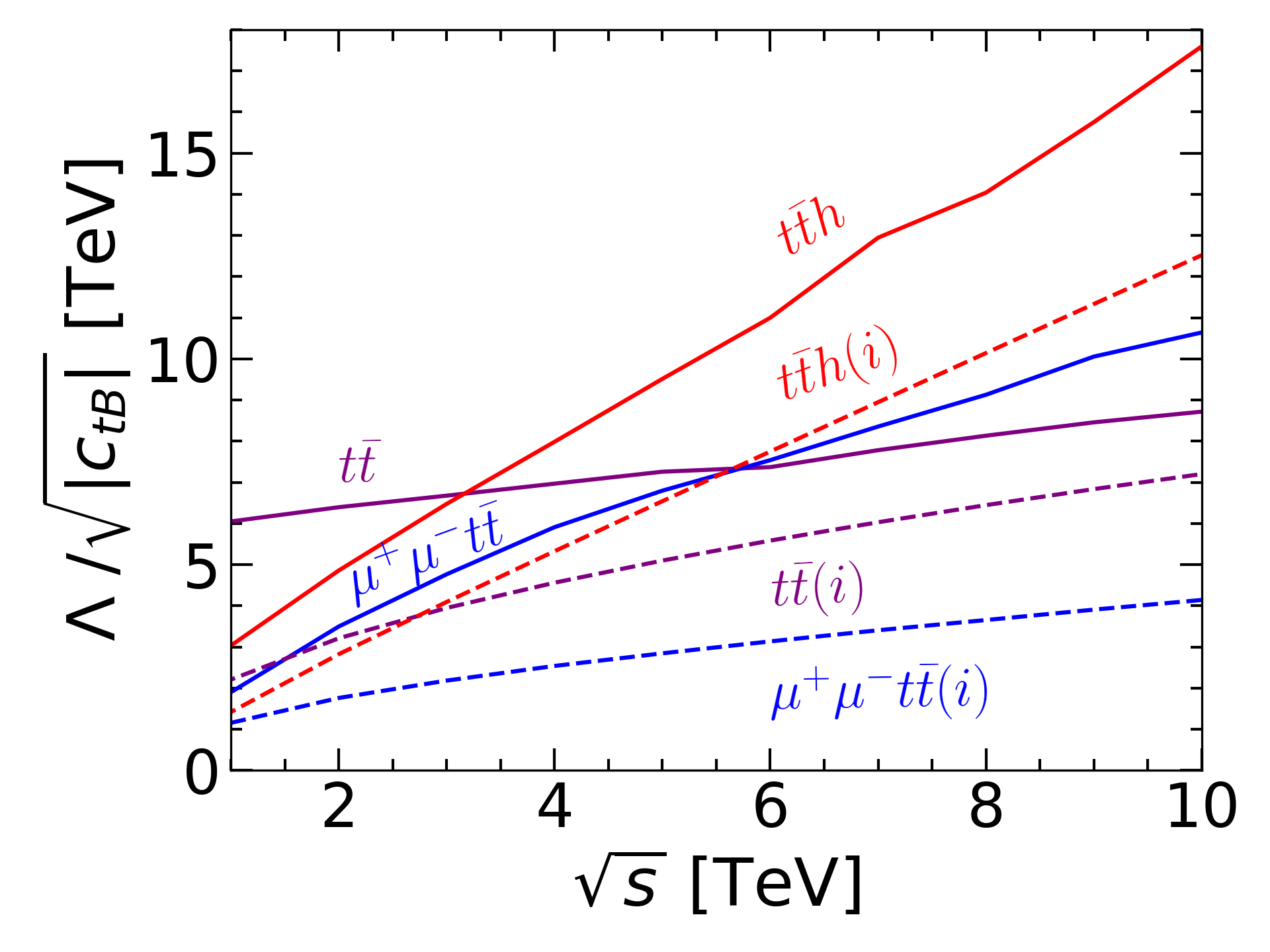}
    \caption{$2\sigma$ sensitivity on $\Lambda/\sqrt{|c_{tB}|}$ versus c.m.~energy from the processes  $\mu^+ \mu^- \to t \bar t,\ t \bar t h$ and $\mu^+ \mu^- \to \mu^+ \mu^- t \bar t$, assuming the integrated luminosity given in Eq.~(\ref{eq:lumi}). The label $(i)$ on the dashed curves indicates an imaginary $c_{tB}$. 
    }
\label{fig:mumusqsq}
\end{figure}

Our final plot Fig.~\ref{fig:mumusqsq} shows the 
$2\sigma$ sensitivity on $\Lambda/\sqrt{|c_{tB}|}$ versus c.m.~energy from the processes  $\mu^+ \mu^- \to t \bar t,\ t \bar t h$ and $\mu^+ \mu^- \to \mu^+ \mu^- t \bar t$. The label $(i)$ on the legends indicates an imaginary $c_{tB}$. For the luminosity, we adopt the desirable scaling relation \cite{Delahaye:2019omf,Aime:2022flm} 
\begin{eqnarray}
\mathcal{L}=  100\ {\rm fb}^{-1} ({\sqrt{s} \over  \text{1 TeV}})^2 .
\label{eq:lumi}
\end{eqnarray}
Once again, we see the complementarity among the processes, with $\mu^+ \mu^- \to t \bar t$ most sensitive at low energies near the threshold, and  $\mu^+ \mu^- \to t \bar t h$ increasing sensitivity at higher energies, and $\mu^+ \mu^- \to \mu^+ \mu^- t \bar t $ contributing moderately. 

\section{Summary and Conclusions}
\label{sec:sum}
Electroweak dipole operators are of great interest to study at high energy colliders. They may be sensitive to new physics in the electroweak sector as the precision measurements have guided us in the past. They are particularly relevant to interactions with heavier fermions because of their unique chiral structure. The coupling strength can be enhanced in dynamical symmetry breaking and composite Higgs models. 
In this paper, we studied the phenomenology of the top quark electroweak dipole operators 
Eq.~(\ref{eq:operator}) at a multi-TeV muon collider. This is motivated by the fact that the dipole operators involving the top quark may be the most significant one due to its heavy mass. As shown in Table \ref{tab:wilsoncoeff} from naive power counting arguments based on some typical strongly interacting models, the Wilson coefficients may be as large as $c_{tB},\ c_{tW} \approx g_* \sim 4\pi$.
Besides, the operators cannot be effectively explored at hadron colliders such as the LHC due to the overwhelming production of $t\bar t$ from QCD and the large SM background. 

We summarized the constraints on $a_V$ and $a_A$ from the current existing experiments and the future collider expectations in the literature and translated them to the lower bounds on the Wilson coefficients of our interests $c_{tB}$, $c_{tW}$ and $\Lambda$, shown in Table \ref{tab:constrain_av}. Some of the bounds from future colliders are expected to be quite complementary to our findings at a muon collider. 

In our study, we first showed the total cross sections for $t\bar t X$ production versus $\mu^+\mu^-$ c.m.~energy in Fig.~\ref{fig:mumu_Ecm}. The effects from the initial state radiation (ISR) have been included throughout the analyses. The $s$-channel processes $\mu^+ \mu^- \to t \bar t,\ t \bar t h$ fall as $1/s$, while the VBF fusions processes 
$\mu^+ \mu^- \to \mu^+ \mu^- t \bar t,\  \nu \bar \nu t \bar t$ grow with energy logarithmically. The solid curves include the electroweak dipole operator contributions for the case $c_{tB}=1$ and $\Lambda = 2 \sqrt s$. All the processes but the $WW$ fusion 
$\mu^+ \mu^- \to  \nu \bar \nu t \bar t$ exhibit certain sensitivity to the new physics. 

The contributions and the scaling behavior for the leading processes $\mu^+ \mu^- \to t \bar t $ and $\mu^+ \mu^- \to t \bar t h$ from the electroweak dipole operators are tabulated in Table \ref{tab:Modelmuon_xsec}. 
Because of the chirality-flipping interactions, the leading interference term for the $\mu^+ \mu^- \to t \bar t $ decreases as $m_t^2/E^2$. The contributions to the $\mu^+ \mu^- \to t \bar t h$ processes, on the other hand, grow with energy as $E^2/\Lambda^2$, resulting in a better sensitivity at a high collision energy. We calculated the SM background and the individual new physics contributions to those processes for $\sqrt s=1,\ 3,\ 10$ TeV, shown in Table \ref{tab:Modelmu_tree_xsec10}. We set the cutoff to be $\Lambda = 2\sqrt s$ and adopted relatively large $c_{tB}$ as guided in Eq.~(\ref{eq:c}). 
We also presented in Fig.~\ref{fig:dist_mumu2tt} the kinematic distributions and showed the differences between the $s$-channel and the VBF processes, that can serve to separate those channels observationally.

Our results are summarized in Tables \ref{tab:Modelmuon_sum} and \ref{tab:Modelmuon2} for real Wilson coefficients $c_{tB}, \ c_{tW}$ and an imaginary $c_{tB}$, respectively. We extracted the $2\sigma$ sensitivity to the model parameters of the electroweak dipole operators. As discussed earlier, the $t \bar t$ channel is more sensitive to the dipole operator effect near the threshold, while the $t \bar t h$ channel is more sensitive at higher energies. The $\mu^+\mu^- t\bar t$ channel contributes moderately.
 We see that the sensitivity to an imaginary coefficient for an electric dipole is not as strong as the magnetic dipoles, due to the absence of the leading interference term. Finally, we showed the sensitivity for all the processes on $\Lambda/\sqrt{|c_{tB}|}$ versus the collider c.m.~energy in Fig.~\ref{fig:mumusqsq},  assuming the luminosity relation Eq.~(\ref{eq:lumi}). 

 In conclusion, a high energy $\mu^+\mu^-$ collider may offer a unique opportunity to probe the electroweak dipole operators associated with the top quark, beyond the coverage by the LHC and future hadron colliders. The significant sensitivity is achieved by the leading channel $\mu^+ \mu^- \to t \bar t$ near the threshold, and substantially improved by the novel channel $\mu^+ \mu^- \to t \bar t h$ at high energies. We may be able to reach a new physics scale well above the collider energy for moderate couplings of the Wilson coefficients. We may be able to probe a new physics scale well above the collider energy for moderate couplings of the Wilson coefficients. Although we performed our analyses with a high energy $\mu^+\mu^-$ collider, our main conclusions are also applicable to a high energy $e^+e^-$ collider, such as the ILC, FCC-ee/CEPC, and CLIC.


\acknowledgements
This work was supported in part by the U.S.~Department of Energy under grant Nos.~DE-SC0007914 and in part by the Pitt PACC.  TH would like to thank the Aspen Center for Physics, where part of this work is complete, which is supported by the National Science Foundation (NSF) grant PHY-1607611, and the CERN TH Department for hospitality during the final stage of the project. 

\bibliographystyle{utphys}
\bibliography{ref.bib}
\end{document}